\documentclass[useAMS,usenatbib]{mn2e}
\usepackage{float,graphics,epsfig,array,amsmath,amssymb,times}

\title[Bayesian galaxy shape measurement]
{Bayesian Galaxy Shape Measurement for Weak Lensing Surveys -I. 
Methodology and a Fast Fitting Algorithm}
\author[L. Miller et al. ]
{L. Miller$^1$, 
T.D. Kitching$^1$, 
C. Heymans$^{2,3}$, 
A.F. Heavens$^4$,
L. Van Waerbeke$^2$.\\
\\
$^1$Dept. of Physics, Oxford University, Keble Road, Oxford OX1 3RH, U.K.\\
$^2$Dept. of Physics and Astronomy,
University of British Columbia, 6224 Agricultural Road,
Vancouver V6T 1Z1, BC, Canada.\\
$^3$Institut d'Astrophysique de Paris, UMR7095 CNRS, 98 bis bd Arago, 75014 Paris, France.\\
$^4$SUPA\thanks{The Scottish Universities Physics Alliance}, Institute for Astronomy, University of Edinburgh,
Blackford Hill, Edinburgh EH9 3HJ, U.K.\\
}

\begin{document}

\pagerange{\pageref{firstpage}--\pageref{lastpage}} \pubyear{2007}

\maketitle

\label{firstpage}

\begin{abstract}
The principles of measuring the shapes of galaxies by a model-fitting approach are
discussed in the context of shape-measurement for surveys of weak gravitational lensing.
It is argued that such an approach should be optimal, allowing measurement with maximal
signal-to-noise, coupled with estimation of measurement errors.  The distinction between
likelihood-based and Bayesian methods is discussed.  Systematic biases in the Bayesian
method may be evaluated as part of the fitting process, and overall such an approach
should yield unbiased shear estimation without requiring external calibration from simulations.
The principal disadvantage of model-fitting for large surveys is the computational
time required, but here an algorithm is presented that enables large surveys to be 
analysed in feasible computation times.  The method and algorithm is tested on simulated
galaxies from the Shear TEsting Program (STEP).
\end{abstract}


\section{Introduction}

Measurement of the effects of weak gravitational lensing has become
a key technique in the arsenal of methods used to measure the 
distribution of matter, both associated with individual objects
such as galaxy clusters or individual galaxies, 
and on large-scales through the measurement of
`cosmic shear'.  A key advantage of such measurement is that
it directly measures the total matter distribution, 
generally dominated by the dark matter component,
which may then be related directly to theory without needing to understand the
uncertain effects of the physics of baryons in galaxies,
provided one avoids the highly nonlinear regime 
(\citealt{white}, \citealt{zhan}, \citealt{jing}).
Through the use of photometric redshifts, three-dimensional analyses 
\citep{Hu,BaconTaylor,Heavens} can be used to further measure
both the cosmological growth of structure and the values of
cosmological parameters \citep{massey07a,Heavens06,Taylor07,Kitching}.  
Until recently such surveys have been of limited size, but even so the
results obtained provided useful constraints on cosmological
parameters and an important test of the 
values deduced from other methods.  One long-standing puzzle
has been that the range of values for the power-spectrum normalisation
parameter $\sigma_8$ found by weak lensing analyses
has tended to be higher than found by 
some other methods (see the discussion in \citealt{spergel07}),
an effect that persists at some level in the latest studies. 
For the best fit 3-year WMAP value of 
the matter density parameter $\Omega_0 = 0.24$,
the 3D analysis of \citet{massey07b} finds the value 
$\sigma_8 = 0.96^{+.09}_{-.07}$, and the 2D analysis of 
\citet{benjamin07} finds $\sigma_8 = 0.84 \pm 0.07$.  
These results can be  
compared with the 3-year WMAP value $\sigma_8 = 0.76 \pm 0.05$ 
\citep{spergel07}.

Measurement of the effect of weak gravitational lensing 
requires the statistical
analysis of large samples and is sensitive to any
systematic errors in measured quantities.  Possible systematic
errors in lensing signals introduced by uncertainty in photometric redshifts
has been discussed by \citet*{edmondson}.  Another fundamental concern with
the method is whether the shapes of galaxies, that are used to deduce the
signal, may be measured in an unbiased manner.  The problem of shape
measurement in optical imaging data is that galaxy images are convolved
with a possibly-varying point-spread function (PSF) which must be accurately
corrected for when deducing galaxy shape.  
Convolution with the PSF tends to make galaxy images appear rounder
(for reasonably circularly symmetric PSFs) whereas 
addition of photon shot noise 
has the systematic effect of tending to make round galaxies 
appear less round.  
These two observational effects thus
tend to work in opposite senses, and are independent of each other, so that 
both accurate PSF correction and calibration to remove the effects of noise
on shape are required.  Following the seminal paper by \citet*{ksb} there have been
many suggestions for possible measurement processes, that are discussed by
\citet{heymans} and \citet{massey07b} as part of the `Shear TEsting Program' (STEP).
Those papers discuss 18 published methods for shear measurement.
The existence of so many suggested methods implies that no consensus has yet emerged
on the best way to measure weak lensing signals, and therefore naturally leads us
to ask whether there might in fact be one method that may be regarded as being optimal.
In this paper we investigate whether a model-fitting approach to galaxy shape
measurement can both achieve this aim of optimal measurement and also be constructed
such that it is computationally feasible for large surveys.

In the following we shall suppose that galaxies may be characterised by a measurement of
their ellipticity {\bf e} and that a weak lensing signal, such as cosmic shear, may be 
inferred either from the mean ellipticity or from some form of cross-correlation of
the ellipticities of different galaxies.

We can stipulate a number of requirements that a weak lensing measurement technique 
should satisfy.
\begin{enumerate}
\setlength\itemsep{0em}
\item Optimal measurement of lensing signal, in the sense of maximum signal-to-noise.
\item Unbiased measurement of lensing signal.
\item Ability to calculate the statistical uncertainties of the measurement.
\end{enumerate}
A standard approach that in principle allows us to meet these criteria is that of model-fitting, which is the
method discussed in this paper.  We first discuss some general principles, including
whether we should use a frequentist or Bayesian approach and how shear may be measured
in an unbiased way from a Bayesian posterior probability distribution.  However, the principal
disadvantage of a model-fitting approach is that it might be computationally prohibitive
for very large surveys.  In section\,\ref{lensfitsection} we discuss a novel 
galaxy shape model fitting algorithm that allows good estimation of the likelihood surface
in a usefully short computational time.  
We also discuss the evaluation of shear sensitivity within the Bayesian framework,
that allows individual galaxy contributions to be assessed and unbiased estimation
of shear to be made, fulfilling the second criterion above.
Some initial results and further considerations
are then discussed.  More detailed results from applying the algorithm to the STEP
simulations are given in a companion paper (Kitching et al., in preparation).

\section{A model-fitting approach to shape measurement}

\subsection{General considerations}
The basic rationale for fitting a model of a galaxy's surface brightness
distribution is that, if the family of models is a good representation
of the true surface brightness profile, the highest possible signal-to-noise 
of the resulting parameters should be obtained. When
model and data agree the model encapsulates the full information content
of the data.  
Although this has been recognised previously in weak lensing shape
measurement \citep{BernsteinJarvis},
no implementation of weak lensing shape measurement methods published to date
has this property, because the methods usually adopt some simplification
of the surface brightness profile, such as assuming that second moments
entirely characterise the profile (e.g. \citealt*{tyson},\citealt{ksb}) or equivalently
assuming Gaussian profiles or weights \citep[e.g.][]{bridle,kuijken99, 
BernsteinJarvis,bardeau05}.  
Model-fitting has been used for some time for detailed determination of
galaxy surface brightness profiles and shapes \citep[e.g.][]{peng}.
\citet{kuijken99} proposed model-fitting to averaged galaxy images specifically
for weak lensing measurement, 
and \citet{bridle} proposed a method of measuring shear by fitting galaxies
and PSFs with multiple Gaussian components.
The latter method has been applied to surveys of weak lensing around galaxy clusters
by \citet{bardeau05}, \citet{bardeau07} and \citet{kneib}, among others. A
Monte-Carlo method is used to find best-fitting galaxy model parameters for each
individual galaxy, where Gaussian surface brightness
profiles, or combinations of two Gaussian profiles, are assumed for both galaxy
and PSF.  Shear measurement and the computational time required
for that model-fitting method has been evaluated by \citet{heymans}.
Recently, sets of
basis functions known as `shapelets' have been used to describe surface
brightness profiles 
(\citealt{refregier, RefregierBacon} and the related work of \citealt{BernsteinJarvis})
but there is no requirement for
the individual shapelet functions to match real galaxy profiles.  Moving to
a pure model-fitting approach allows us to choose whichever brightness profiles
we like, and for galaxies it clearly makes most sense to choose either
exponential or de Vaucouleurs surface brightness profiles.  Naturally, the above
statements are qualitative, we don't know {\em how much} improvement one
obtains by fitting a profile that is closer to the actual profile, but the
principle at least is a sound one, that we expect to satisfy the first of
our criteria from the Introduction.  

Either frequentist model-fitting, based on determining the likelihood function,
or Bayesian model-fitting that determines the posterior probability distribution
of model parameters, allow error estimates to be made, satisfying the third of
our criteria.  This is not the case for early versions of weak lensing shear estimators,
although error estimates have been made in some recent methods 
\citep{BernsteinJarvis, kuijken06, bridle, bardeau05}.

Finally, we should address the question of whether a method can be determined
to be unbiased.  This is a serious issue for weak lensing studies, where the
signal is so small that even a small systematic bias can have a devastating effect.
Evaluation of existing methods by STEP demonstrate that they are indeed biased,
with significant magnitude-dependent biases that need to be corrected empirically
from comparison with simulations \citep{heymans,massey07b}.  We discuss in the next 
section why in principle a Bayesian method should be unbiased provided a correct
choice of prior is made, but note that realistic implementations result in a
quantifiable bias that may be corrected for.

\subsection{Bayesian estimation of the sample ellipticity distribution}
\label{sampledistribution}

We have previously mentioned the problem of shape measurement, that not only
is the shape changed by convolution with the PSF, but also noise biases
the measured shape, and in general tends to make nearly-circular objects
systematically appear more elliptical.  We discuss in this section how a Bayesian method
may be formulated that precisely corrects for this phenomenon, provided we
make a correct choice of prior.

Consider a set of observations of $N$ galaxies that yields the 
surface brightness distribution
for each galaxy denoted by a vector of pixel values $\bmath{y}$.  
The shape of each galaxy may be characterised by its two-component ellipticity
$\bmath{e}$: the particular definition we choose for $\bmath{e}$ in 
this paper is given in section\,\ref{shearestimation} but what follows below applies
to any shape estimator that we may choose.
If the sample of galaxies has a
probability distribution of intrinsic ellipticities (i.e. the value of ellipticity that
would be measured by the observer in the absence of degradation by the PSF or
by noise) $f(\bmath{e})$, then the probability distribution of $\bmath{y}$
is
\[
n(\bmath{y}) = \int f(\bmath{e}) \epsilon(\bmath{y}|\bmath{e}) \mathrm{d}\bmath{e},
\]
where $\epsilon(\bmath{y}|\bmath{e})$ is the probability distribution 
for $\bmath{y}$ given $\bmath{e}$.

For each of these galaxies we can generate a Bayesian posterior probability
distribution for its ellipticity
\[
p_i(\bmath{e} | \bmath{y}_i) = 
\frac
{\mathcal{P}\left (\bmath{e}\right ) \mathcal{L}\left (\bmath{y}_i | \bmath{e}\right )}
{\int\mathcal{P}\left (\bmath{e'}\right ) \mathcal{L}\left (\bmath{y}_i | \bmath{e'}\right ) \mathrm{d}\bmath{e'}}
\]
where $\mathcal{P}\left (\bmath{e}\right )$ is the ellipticity prior
probability distribution and
$\mathcal{L}\left (\bmath{y}_i | \bmath{e}\right )$
is the likelihood of obtaining the $i^{th}$ set of data values $\bmath{y}_i$
given an intrinsic ellipticity $\bmath{e}$.

We would hope that the true distribution of intrinsic ellipticities can
be obtained from the data by considering the summation over the data:
\begin{eqnarray}
\nonumber
\lefteqn{
\langle
\frac{1}{N}
\sum_i p_i(\bmath{e} | \bmath{y}_i) 
\rangle
= } \\
\nonumber
& & 
\int \mathrm{d}\bmath{y} \frac
{\mathcal{P}\left (\bmath{e}\right ) \mathcal{L}\left (\bmath{y} | \bmath{e}\right )}
{\int\mathcal{P}\left (\bmath{e'}\right ) \mathcal{L}\left (\bmath{y} | \bmath{e'}\right ) \mathrm{d}\bmath{e'}}
\int f(\bmath{e''}) \epsilon(\bmath{y}|\bmath{e''}) \mathrm{d}\bmath{e''}
\end{eqnarray}
where on the RHS we are integrating over the probability distributions to
obtain the expectation value of the
summed posterior probability distribution for the sample.
We can see that this will be achieved if both
$\epsilon(\bmath{y} | \bmath{e}) = \mathcal{L}\left (\bmath{y} | \bmath{e}\right )$
and 
$\mathcal{P}\left (\bmath{e}\right ) = f\left (\bmath{e}\right )$, assuming the
likelihood is normalised,
$\int \mathcal{L}\left (\bmath{y} | \bmath{e}\right) \mathrm{d}\bmath{y} = 1$,
from which we obtain
\[
\langle
\frac{1}{N}
\sum_i p_i(\bmath{e} | \bmath{y}) 
\rangle
= \mathcal{P}\left (\bmath{e} \right )
= f(\bmath{e}).
\]

The strength of this result is that we can in principle recover statistically knowledge
of the intrinsic distribution of shapes independently of assumptions about
the shapes of the likelihood surfaces: in particular the likelihood surfaces
for ellipticity measurement must be non-Gaussian, being bounded at
$|e|<1$, but this has no effect on the results we expect.
This result parallels the analogous result discussed by \citet{edmondson} for
the case of Bayesian photometric redshift estimation.  It says that we must know
the mechanism by which data values are generated in order to construct the
likelihood function, and that we must know the expected distribution of intrinsic
ellipticities, in which case the summed posterior probability distribution
will recover that intrinsic distribution.  It might be thought that a Bayesian
approach then has a non-useful requirement, that we need to know the answer
before we start, but the point of course is that with the correct choice of
prior we then expect the posterior probability distribution for each {\em individual}
galaxy to yield an unbiased estimate of ellipticity, and those sets of individual
posterior probability distributions may then be used to infer the spatially
varying shear arising from gravitational lensing.  We discuss in section\,\ref{sec:prior}
one possible method for creating the correct prior.

\subsection{Frequentist or Bayesian measurement?}
So far the framework has been described in a purely Bayesian context, but we can also
ask whether there is a frequentist equivalent of the above formalism: can weak lensing
shear be measured using likelihood functions alone?  Conversely, are there any disadvantages
to using a Bayesian method? 

\begin{figure}
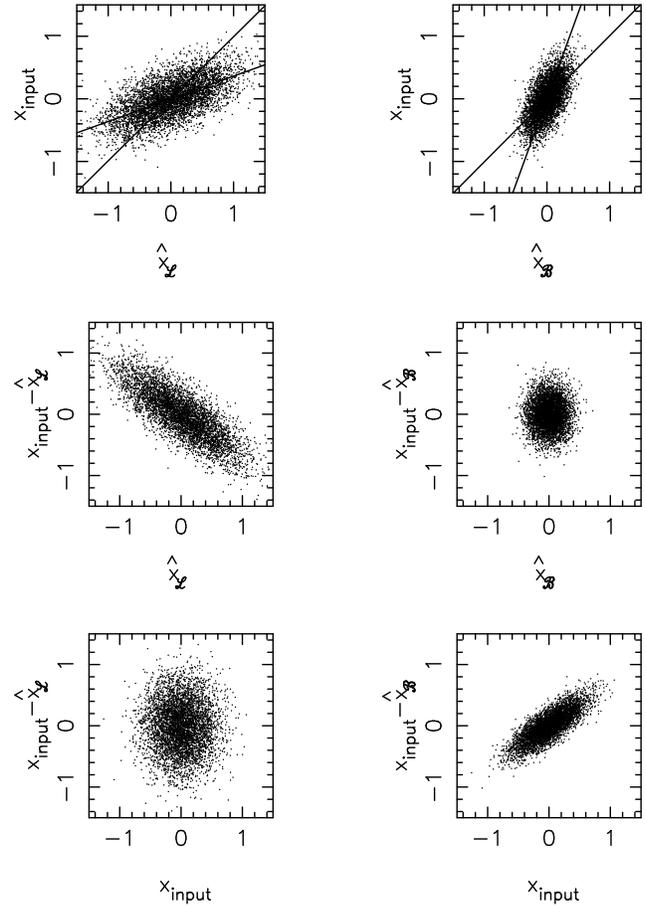

\resizebox{84mm}{!}{
\rotatebox{-90}{
\includegraphics{fig1a.ps}}}

\vspace{5mm}

\resizebox{84mm}{!}{
\rotatebox{-90}{
\includegraphics{fig1b.ps}}}

\vspace{5mm}

\resizebox{84mm}{!}{
\rotatebox{-90}{
\includegraphics{fig1c.ps}}}
\caption{
Illustration of the properties of 
an ideal likelihood estimator $\hat{x}_{\mathcal{L}}$ (left) and
ideal Bayesian estimator $\hat{x}_{\mathcal{B}}$ (right) 
for the Gaussian example described in the text.  
The top pair of graphs show the correlation between the input and deduced
values.  
Two regression lines are shown on each, one being the regression of input on estimated
value, the other being the regression of estimated on input value.
The next two pairs show the distribution of the difference between input and estimated
values compared with either the estimated values
(centre) or the input values (bottom).  Note the graph x-axes differ between the
centre and bottom panels.
For a given input value, the likelihood estimator yields an unbiased estimate
(regression slope unity) whereas the Bayesian estimator appears biased (regression slope
2.75).  However, for a given estimated value, the likelihood estimator is biased
(regression slope 0.36) and the Bayesian estimator is unbiased.  The Bayesian estimator
returns the best estimate of the input value for a given measurement.}
\label{example}
\end{figure}

It is important to recognise that likelihood and Bayesian estimators measure different
things.  We can illustrate this by considering a sample of galaxies (say) with some
intrinsic property $x$ that we wish to determine from fitting to some measurements $y$.
We shall look at the results obtained with either a Bayesian estimator,
$\hat{{x}}_{\mathcal{B}} = \int {x} p(x|y) \mathrm{d}{x}$,
or a likelihood estimator,
$\hat{{x}}_{\mathcal{L}} = \int {x} \mathcal{L}(y|x) \mathrm{d}{x}$
(the general considerations discussed here apply also to maximum likelihood estimators).
Suppose the intrinsic distribution of $x$
has a normal distribution of variance $a^2$, and that 
for each $x$ drawn from this distribution
the measurement process causes a
normally distributed uncertainty of variance $b^2$.  Fig.\,\ref{example} shows the
results obtained in a Monte-Carlo realisation for the illustrative case $a=0.3$, $b=0.4$.

The likelihood estimator is based on the function $\mathcal{L}(y|x)$ 
and in Fig.\,\ref{example}
is unbiased in the regression of input on estimated values.  The Bayesian estimator
is based on the function $p(x|y)$ and is unbiased in the regression of estimated on
input values: that is, for a given set of measurements the Bayesian estimator yields the
best estimate of the input values.  The likelihood estimator yields a distribution
of measured values that is broader than the intrinsic distribution.  Notice however that
the Bayesian estimator yields a distribution that is {\em narrower}, despite the result 
of section\,\ref{sampledistribution} that the summed posterior probability distribution
yields the intrinsic distribution if the correct prior is chosen.  This apparent paradox
is resolved by realising that each of the individual estimated values is associated with
its own posterior probability distribution, so that the sum of the distributions is broader
than the distribution of expectation values.  While this might seem undesirable, this is
inevitable in any noisy measurement process.

In the case of ellipticity shape measurement, we expect there to be other reasons why
a likelihood estimator might be biased, in particular
at large $\bmath{e}$ values or for low signal-to-noise, where the
boundary $|\bmath{e}| < 1$ renders the likelihood function asymmetric and highly non-Gaussian.
With no prior, the strong degeneracy between
size and ellipticity in the likelihood fitting can create regions of high likelihood at extreme
values of ellipticity, and given the hard bound $|\bmath{e}| < 1$ such an estimator cannot be
unbiased.

In the frequentist approach, it is also possible to estimate errors for the individual
galaxies, and this is important to establish the contribution to the signal from each
galaxy.  As in the Bayesian approach, we expect the information on shear to decrease as the
signal-to-noise (S/N) decreases: even with an unbiased estimator $\hat{\bmath{e}}$ it would be
important to quantify this effect and allow for it in the shear estimation.

A final consideration is that in weak lensing surveys we are not simply interested in
measuring the shapes of individual galaxies, but rather in measuring the systematic 
lensing shear in a sample.  Usually this is measured from the mean ellipticity: thus it
may be possible to have an ellipticity estimator that is biased but where the shear estimate
from a sample is unbiased, or {\em vice versa}.  If any bias were isotropic, corresponding
to a bias in the value of $|\bmath{e}|$ but not in orientation, then we might hope that
the bias would average out.  However, even in this case we should not {\em assume} that the
shear estimator $\langle\bmath{e}\rangle$ is unbiased, since the likelihood functions
for $\bmath{e}$ measurement must be non-Gaussian and $\bmath{e}$-dependent, any shift
$\bmath{g}$ in the distribution of $\bmath{e}$ caused by lensing would lead to bias in
the estimated shear.

However, even the Bayesian method is not immune to the problem of bias, particularly
in a realistic implementation of the Bayesian method where
we are forced to assume a zero-shear prior, as discussed below.
But the bias can be quantified and the method
provides a self-contained framework within which we can work out all the 
required quantities. 
This is the framework that we return to in the remainder of this
paper.

\subsection{Bayesian shear estimation and the shear sensitivity}
\label{shearestimation}
Following \citet{heymans} we assume observed galaxy ellipticity $\bmath{e}$ is
related to the intrinsic galaxy ellipticity $\bmath{e^s}$ in the weak lensing regime
via:
\[
\bmath{e} = \frac{\bmath{e^s} + \bmath{g}}{1 + \bmath{g^{\star}e^s}}
\]
from \citet{SchrammKayser,SeitzSchneider}, where $\bmath{e}$ is represented as a complex variable and
$\bmath{g}$, $\bmath{g^{\star}}$ are the reduced shear and its complex conjugate
respectively.  $\bmath{e}$ is defined in terms of the major and minor axes and orientation
$a, b, \theta$ respectively, as $\bmath{e} = (a-b)/(a+b)\exp(2i\theta)$.
In this formalism, we expect
\begin{equation}
\langle \bmath{e} \rangle = \bmath{g}
\label{eqn:weakshear}
\end{equation}
for an unbiased sample where $\langle \bmath{e^s} \rangle = 0$,
and so $\langle \bmath{e} \rangle$ 
for a sample of galaxies is adopted as our estimator
of $\bmath{g}$. Note that this result differs from the other commonly used
formalism where ellipticity is instead defined as 
$\bmath{e} = (a^2-b^2)/(a^2+b^2)\exp(2i\theta)$.

For a population of galaxies,
$
\langle\bmath{e}\rangle = \int \bmath{e} f(\bmath{e}) \mathrm{d}\bmath{e}
$
where $f(\bmath{e})$ is the ellipticity probability distribution for the sample.  But
in the Bayesian formalism we can write a similar expression for an individual galaxy 
if we know its Bayesian posterior probability distribution,
$
\langle\bmath{e}\rangle_i = \int \bmath{e} p(\bmath{e}|\bmath{y}_i) \mathrm{d}\bmath{e}.
$
Hence for a sample of $N$ galaxies we can evaluate the sample mean as
\[
\langle\bmath{e}\rangle = \frac{1}{N} \sum_i \int \bmath{e} p_i(\bmath{e}|\bmath{y}_i) 
\mathrm{d}\bmath{e} 
= \frac{1}{N} \int \bmath{e} \sum_i p_i(\bmath{e}|\bmath{y}_i) \mathrm{d}\bmath{e}.
\]
In practice we shall use the first of these expressions, as estimation of
ellipticities for individual galaxies allows error estimates to be made
for each galaxy, and its contribution to the signal to be evaluated.

However, in measuring shear we cannot know in advance the correct prior to apply,
even if we know the intrinsic unsheared ellipticity prior distribution, because 
the amount of shear varies over the sky in a way that we are attempting to measure.
We must therefore use a prior that contains zero shear.  The effect of this is that
as signal-to-noise decreases, the measured ellipticity distribution tends to the
prior, and in the limit of zero signal-to-noise no shear signal is recoverable.
This is precisely what we should expect of course: no measurement method can extract
a measured shear value from data with zero signal-to-noise, and a Bayesian method
is no different in that respect.  A Bayesian method does however allow us to
estimate the magnitude of this effect for each individual galaxy.  Consider the Bayesian
estimate of ellipticity $\langle \bmath{e} \rangle_i$, defined above,
that is measured for the $i^{\mathrm{th}}$ galaxy, and
express its dependence on each component of shear $\bmath{g}$ as a Taylor series.  For
component $e_1$,
\begin{equation}
\langle e_1 \rangle_i
\simeq
{e_1^s}_i +
g_1 \partial\langle e_1\rangle_i/\partial g_1 +
g_2 \partial\langle e_1\rangle_i/\partial g_2 +
\ldots
\label{eqn:taylorseries}
\end{equation}
and similarly for component $e_2$,
where numeric subscripts indicate the components of $\bmath{e}$ and $\bmath{g}$.
In the weak lensing limit the cross-terms vanish.  
If we sum over 
$N$ galaxies in an unbiased sample we find
\[
\sum_i^N \langle e_1 \rangle_i \simeq
g_1 \sum_i^N \partial\langle e_1\rangle_i/\partial g_1.
\]
We may optionally multiply both sides in equation\,\ref{eqn:taylorseries}
by a statistical weight for each galaxy,
$w_i$. Provided $w_i$
is uncorrelated with $\bmath{e}^s_i$
we may then define a weighted estimate of shear for the sample:
\begin{equation}
\hat{g_\mu} \equiv \frac{\sum_i^N w_i \langle e_\mu \rangle_i}
{\sum_i^N w_i \partial\langle e_\mu \rangle_i/\partial g_\mu }
\end{equation}
for $\mu = 1,2$.
We shall call $\partial\langle e_\mu \rangle_i/\partial g_\mu$ 
the {\em shear sensitivity}. 
It is a measure of how much each Bayesian estimate is biased by the use
of the zero-shear prior, and it takes values in the range
$0 < \partial\langle e_\mu \rangle_i/\partial g_\mu \le 1$,
where the lower bound is expected in the limit of zero signal-to-noise. The
upper bound would be attained in the case of ideal measurement at high signal-to-noise, 
where we expect no bias: in this case the Bayesian measure is a 
good measure of the true ellipticity, regardless of which prior is assumed,
and differentiating equation\,\ref{eqn:weakshear} yields unity for
the shear sensitivity.
The weights $w_i$ could in principle be tuned for 
optimal signal-to-noise in the measurement(s) being made, 
such as the values of cosmological parameters.  
Care should be taken that any
weights that are a function of ellipticity do not introduce bias
into shear measurement.
Since the shear sensitivity is effectively a measure of how much information
about the effect of lensing is carried by each galaxy, 
the weights should themselves also include a dependence on
$\partial\langle e_\mu \rangle_i/\partial g_\mu$, as well as on the 
measurement error and on
the redshift-dependent cosmological effect of lensing on each galaxy.

The shear sensitivity
may be estimated for each galaxy and for the survey as
a whole without recourse to external calibration from simulations,
as described below.
\citet{ksb,Luppino,kaiser00} and \citet{BernsteinJarvis}
have emphasised the utility of knowing the shear `polarizability' or `responsivity'
for individual galaxies, as this not only allows
accurate optimised shear measurement but also allows future surveys to be 
planned and optimised.

The estimator $\hat{g}_\mu$ is appropriate for a survey where the shear is uniform over
some region (and this is assumed in the STEP simulations discussed below),
but in the more general case we instead infer the shear correlation function
or some related quantity such as shear variance from a measurement such as
$\langle\bmath{e}_i\bmath{e}_j\rangle$.  In this case
we can compute the analogous estimator
\[
\langle\widehat{g_\mu g_\mu}\rangle = 
\frac{\sum_{i,j} w_i w_j \langle e_\mu \rangle_i \langle e_\mu \rangle_j}
{\sum_{i,j} w_i w_j 
\left(\partial\langle e_\mu \rangle_i/\partial g_\mu \right)
\left(\partial\langle e_\mu \rangle_j/\partial g_\mu \right)
}.
\]

We now discuss possible approaches to calculating the shear sensitivity, first
for normal prior and likelihood distributions, then for the more general case
where the shear sensitivity may be evaluated numerically from the measured
likelihood surfaces of individual galaxies.

\subsection{Calculation of shear sensitivity}
\label{sec:shearnormal}
As an illustration of the calculation of shear sensitivity, 
suppose the prior is described by a normal distribution $\mathcal{P}(\bmath{e})$
of variance $a^2$ and $\langle\bmath{e}\rangle=0$, 
and that the likelihood $\mathcal{L}(\bmath{e})$
for a particular galaxy also has a normal distribution of
variance $b^2$ centred on some value $\bmath{e}_0$.  It is straightforward then to
show that the Bayesian posterior probability $p(\bmath{e}|\bmath{y})$ also has a normal
distribution of variance $a^2b^2/(a^2 + b^2)$ and expectation value
$\langle\bmath{e}\rangle = \bmath{e}_0 a^2/(a^2 + b^2)$.  For perfect measurement of
ellipticity ($b^2 \ll a^2$) we expect equation\,\ref{eqn:weakshear} to hold, so
for this galaxy 
$\partial\langle e_\mu \rangle_i/\partial g_\mu = \partial {e_0}_\mu/\partial g_\mu = 1$.
For more noisy measurement, we expect
\[
\frac{\partial\langle e_\mu \rangle_i}{\partial g_\mu} = 
\frac{a^2}{a^2 + b^2} \frac{\partial {e_0}_\mu}{\partial g_\mu} =
\frac{a^2}{a^2 + b^2}.
\]
The shear sensitivity decreases as the measurement error increases.  
The value of the shear sensitivity is also given by the 
inverse of the slope of the regression of the intrinsic ellipticity on estimated ellipticity
illustrated in Fig.\,\ref{example}.

The above illustration indicates that it is straightforward to
calculate the shear sensitivity, however in general it would not be
safe to assume normal distributions: not least because $\bmath{e}$ is
defined such that $|\bmath{e}| < 1$, so when the measurement error
becomes large $\mathcal{L}(\bmath{e})$ cannot be normally distributed.
We discuss here one method of 
calculating the shear sensitivity numerically.  We should
emphasise that this can be done entirely internally to the
fitting process, with no need to calibrate shear sensitivity
externally from simulations.

Consider first the response of the posterior probability distribution to a small amount of shear.
The prior probability does not depend on the shear in our implementation.
Let us assume that applying a weak lensing shear shifts the likelihood function
by some small amount, 
$\mathcal{L}(\bmath{e}-\bmath{e^s})\rightarrow
\mathcal{L}(\bmath{e}-\bmath{e^s}-\bmath{g})$
and expand as a Taylor series:
\[
\mathcal{L}(\bmath{e}-\bmath{e^s}-g_\mu) \simeq 
\mathcal{L}(\bmath{e}-\bmath{e^s}) - g_\mu \frac{\partial\mathcal{L}}{\partial e_\mu} + \ldots.
\]
Then, substituting into 
\[
\langle\bmath{e}\rangle = 
\frac{
\int\bmath{e}\mathcal{P}(\bmath{e})\mathcal{L}(\bmath{e})\mathrm{d}\bmath{e}
}{
\int\mathcal{P}(\bmath{e})\mathcal{L}(\bmath{e})\mathrm{d}\bmath{e}
}
\]
and differentiating with respect to $\bmath{g}$ we find,
\[
\frac{\partial\langle e_\mu \rangle}{\partial g_\mu} \simeq 
\frac{
\int \left(\langle\bmath{e}\rangle-\bmath{e}\right) \mathcal{P}(\bmath{e}) \frac{\partial\mathcal{L}}{\partial e_\mu} \mathrm{d}\bmath{e} 
}
{
\int\mathcal{P}(\bmath{e})\mathcal{L}(\bmath{e})\mathrm{d}\bmath{e}
}
\]
as an estimate of weak lensing shear sensitivity.
This expression is cast in terms of the derivatives of the likelihood surface
multiplied by the prior: it may also be expressed in terms of derivatives
of the prior multiplied by the likelihood:
\[
\frac{\partial\langle e_\mu \rangle}{\partial g_\mu} \simeq 
1 - 
\left[ 
\frac{
\int \left(\langle\bmath{e}\rangle-\bmath{e}\right) 
\mathcal{L}(\bmath{e}) 
\frac{\partial\mathcal{P}}{\partial e_\mu} \mathrm{d}\bmath{e} }
{
\int\mathcal{P}(\bmath{e})\mathcal{L}(\bmath{e})\mathrm{d}\bmath{e}
}
\right].
\]
This may be evaluated numerically
from the posterior probability surface for each galaxy, and is preferred
over the preceding expression in the case where the derivative of the
prior is known analytically.
For the case of normal distributions of $\mathcal{P}(\bmath{e})$ and 
$\mathcal{L}(\bmath{e})$ the expression yields  
the analytic result above.

\section{Fast shape measurement}\label{lensfitsection}
\subsection{The algorithm}
The technique we adopt for measuring $\langle \bmath{e} \rangle$ and its uncertainty
is to fit model galaxy surface brightness profiles to the data for individual
galaxy images.  The simplest galaxy model has six free parameters if the
form of the surface brightness profile is fixed: central surface brightness,
size, ellipticity and celestial position.  The problem of fitting six parameters
to large samples of galaxies is that this could be a time-consuming task,
probably prohibitively so.  However, we can greatly speed up the process if we
can marginalise over any parameters that are not of interest to the weak lensing
measurement. It turns out that {\em for isolated galaxies}
it is straightforward to marginalise over
three of the parameters, central surface brightness and position, if the model
fitting is treated in Fourier space, as described below.  And because there
exist fast Fourier transform algorithms this approach can be done in a short
amount of computational time.

We can start by writing the statistic
\begin{eqnarray}
\nonumber
\chi^2 &=& \sum_i \left[\frac{y_i-Cy^m_i}{\sigma_i}\right]^2\\
\nonumber
 &=& \sum_i \frac{y_i^2}{\sigma_i^2} + A\left(C-B\right)^2 -AB^2,
\end{eqnarray}
where $y_i$ is the data value in pixel $i$, $\sigma_i$ is the statistical
uncertainty of that data value, $y^m_i$ is a model value for that pixel,
$C$ is the model amplitude and where
\[
A = \sum_i\left (\frac{y^m_i}{\sigma_i}\right )^2, \hspace*{1cm}
B = \sum_i\frac{y_i y^m_i}{\sigma_i^2}/\sum_i\left (\frac{y^m_i}{\sigma_i}\right)^2.
\]
We assume the pixel noise is stationary and uncorrelated, which is appropriate for 
shot noise in CCD detectors in the sky-noise limit.  
Bright galaxies may
make a significant contribution to photon shot noise, but in this case
the non-stationarity of the noise makes it not possible to work in Fourier
space.  Hence this algorithm is appropriate for model fitting to faint galaxies
in the sky-limited regime. 
The method can be generalised to the case where the noise is
stationary but correlated between pixels, and the example of radio
interferometer observations is discussed qualitatively in section\,\ref{radio}.

Then if we adopt a prior $\mathcal{P}(C)$ for the model amplitude
we can marginalise the likelihood
$\mathcal{L} = e^{-\chi^2/2}$ over $C$:
\[
\mathcal{L} = e^{-\sum y_i^2/2\sigma^2} e^{AB^2/2} 
\int_{C_{\rm min}}^{C_{\rm max}} e^{-A\left(C-B\right)^2/2} \mathcal{P}(C) \mathrm{d}C.
\]
We shall adopt a uniform prior for $\mathcal{P}(C)$ in the range
$C_{\rm min} \le C \le C_{\rm max}$.
We expect $C>0$, but 
if the galaxy is significantly detected the Gaussian form of the likelihood
causes the value of the prior to become unimportant at both large and small
$C$, and we can simplify the calculation by
allowing $C_{\rm min} \rightarrow -\infty$ and
$C_{\rm max} \rightarrow \infty$, so that
\[
\mathcal{L} \simeq \sqrt{\frac{2\pi}{A}} e^{-\sum y_i^2/2\sigma_i^2} e^{AB^2/2}.
\]
However, although we have eliminated the amplitude $C$, $\mathcal{L}$ still
depends on the model second moment, $A=\sum \left( y^m_i/\sigma_i\right )^2$.  
Thus we need to introduce the model constraint $A=$constant, achieved by 
renormalising each model appropriately. Since for a given dataset
$\sum y_i^2/\sigma_i^2$ is also fixed, we can write
$\mathcal{L} \propto e^{AB^2/2}$ when maximising.

We can also rapidly calculate the marginalisation over galaxy celestial position
if we work in Fourier space, writing
\[
y_i = \sum_k y_k e^{-i\bmath{k.x_i}}, \hspace{1cm}
y^m_i = \sum y^m_k e^{-i\bmath{k.x_i}}.
\]
We can simplify the various summations by assuming that we are dealing with faint
galaxies in weak lensing measurement, such that $\sigma_i$ is dominated by the
background photon shot noise and is constant for all pixels.  And since the model
$y^m_i$ is real, $y^m_k = {y^m_k}^{\star}$ and 
$\sum_i y_i y^m_i = \sum_k y_k {y^m_k}^{\star}$.  
If we introduce a shift $\bmath{X}$ into the model position, the new model becomes
\[
{y^m_i}' = \sum_k y^m_k e^{-i\bmath{k.x_i}} e^{-i\bmath{k.X}}
\]
and
\[
\sum_i y_i {y^m_i}' = \sum_k y_k {y^m_k}^{\star} e^{-i\bmath{k.X}} = h(\bmath{X})
\]
where $h(\bmath{X})$ is the cross-correlation of the data $y_i$ with the model
$y^m_i$.  So the likelihood becomes
\[
\mathcal{L} \propto 
\exp\left[ 
\frac{|h(\bmath{X})|^2}{2\sigma^2 \sum {y^m_i}^2 } 
\right].
\]
To marginalise over $\bmath{X}$ we need to adopt a prior $\mathcal{P}(\bmath{X})$,
but in this case it cannot be uniform as $\mathcal{L}\rightarrow$constant as
$|\bmath{X}|\rightarrow\infty$ and the marginalised likelihood would not be finite.
This problem arises because, no matter how large a pixel value, it always has a finite
chance of being due to random noise, with the true galaxy being positioned elsewhere.
We shall adopt a prior which is centred on some assumed galaxy position that has
been previously estimated and which falls off to zero at large distances: this is
equivalent to assuming that a galaxy does indeed exist somewhere near the location
we have chosen.  We shall assume a prior which is symmetric and centred on the
nominal galaxy position, which for convenience is at the coordinate origin, such as:
\[
\mathcal{P}(\bmath{X})d^2\bmath{X} =
\frac{1}{2\pi b^2} e^{-\left |\bmath{X}\right |^2/2b^2} d^2\bmath{X}.
\]
The process of model fitting is seen from the above to be
one of cross-correlating the data with a model.
Galaxies generally have smooth centrally-concentrated surface brightness 
distributions which are convolved with near-Gaussian PSFs in
an observed image.  The model is also smooth,
centrally concentrated and convolved with the same PSF.  
From the central limit theorem such a cross-correlation
should be well represented by a two-dimensional Gaussian distribution,
\[
h(\bmath{X}) = h_0 
\exp\left[-(\bmath{X}-\bmath{X}_0)\mathcal{C}^{-1}(\bmath{X}-\bmath{X}_0)^{T} \right]
\]
where $\mathcal{C}^{-1}$ is the inverse covariance matrix and 
some shift $\bmath{X}_0$ of the maximum from the origin is allowed.
In what follows we assume circular symmetry for simplicity, although this
assumption may be removed without affecting the final result
(a two-dimensional Gaussian distribution 
may always be transformed to a circularly-symmetric distribution by a 
coordinate transformation). If the
cross-correlation function has the form $h = h_0 \exp [-|\bmath{X}-\bmath{X}_0|^2/s^2]$ then
\[
\mathcal{L} \propto
\frac{1}{2 \pi b^2}
\int_0^\infty \exp \left[
{\beta e^{-\left | \bmath{X}-\bmath{X}_0 \right |^2/s^2}}
\right ]
e^{-\left |\bmath{X}\right |^2/2b^2} \mathrm{d}^2\bmath{X},
\]
where
\[
\beta  = \frac{h_0^2}{2\sigma^2 \sum {y^m_i}^2}.
\]
We could evaluate this by, for example, expanding the first exponential as 
a Taylor series and hence obtaining a series solution for the marginalised
likelihood.  We could also evaluate it purely numerically, but this would require
evaluation of the cross-correlation function on an extremely fine grid
in order to achieve adequate accuracy.
Either of these approaches would be computationally expensive, and an alternative
is to find an approximate value of the integral by writing
\[
{\cal L} \propto \frac{1}{2\pi b^2} \int_0^\infty \left\{\exp\left[\beta e^{-\left |\bmath{X}-\bmath{X}_0\right|^2/s^2}\right]-1+1\right\} e^{-\left |\bmath{X} \right |^2/2b^2} \mathrm{d}^2 \bmath{X}.
\]
If $b \gg s$, 
\[
{\cal L} \propto 1 + \frac{1}{2\pi b^2} 
e^{-\left | \bmath{X}_0 \right |^2/2b^2} 
\int_0^\infty \left\{\exp\left[\beta e^{-\left |\bmath{X}-\bmath{X}_0\right |^2/s^2}\right]-1\right\} \mathrm{d}^2 \bmath{X}
\]
and changing variables to a polar system centred on $\bmath{X}_0$, 
\begin{eqnarray}
\nonumber
{\cal L} & \propto & 1 + \frac{s^2}{b^2} 
e^{-\left | \bmath{X}_0 \right |^2/2b^2} 
\int_0^\infty\, \left\{\exp\left[\beta e^{-r^2}\right]-1\right\} r \mathrm{d}r\\
& \propto &
\frac{s^2 e^{\beta} }{ 2\beta b^2} 
e^{-\left | \bmath{X}_0 \right |^2/2b^2} 
 \qquad \beta\gg 1.
\label{eq:likelihood}
\end{eqnarray}
approximately, where the constant of proportionality has no model dependency
provided $A$ is held invariant
(we could obtain a similar result more exactly if we were to adopt a top-hat
prior for the galaxy position).
If the width $s$, amplitude $h_0$ and centroid $\bmath{X}_0$ of the cross-correlation
function can be measured, the marginalised likelihood may be estimated 
from equation\,\ref{eq:likelihood}.
\begin{footnote}
{It may seem that the requirement for a prior on position may be removed by allowing
$b\rightarrow\infty$. However, this is an
artefact of the approximation.  There is no clear
way of identifying a value for $b$, but it should be set sufficiently small that 
confusion from other nearby galaxies is eliminated.}
\end{footnote}
In the more general case, where the cross-correlation
function is approximated by a bivariate Gaussian with widths $s_1$ and $s_2$
in the two principal directions, then equation\,\ref{eq:likelihood} is modified by
$s^2 \rightarrow s_1 s_2$.

\subsection{Implementation}
Using the above, the likelihood may be estimated for a given set of model parameters,
marginalised over the `uninteresting' position and brightness of the galaxy.
If we choose either an exponential disc or a de Vaucouleurs
model for the surface brightness, the free parameters are the scale-length and
two ellipticity values $\bmath{e}$, or equivalently the scale-length, the axial
ratio and the orientation.  The values of $\bmath{e}$ are restricted to lie in
the range $|\bmath{e}|<1$, and for faint galaxies the probability $p(\bmath{e})$
is broad, making a grid search in $\bmath{e}$ an easy and not-too-expensive approach.
The resulting likelihood may be numerically marginalised over the galaxy scale-length,
which is also `uninteresting' for weak lensing measurement,
to obtain a likelihood surface that is a function of ellipticity alone.

To evaluate the cross-correlation function $h(\bmath{X})$ (returning now to the 
general case where we do not assume circular symmetry) we can use the fast Fourier
transform method, which proceeds as follows:
\begin{enumerate}
\item Generate a series of 2D galaxy surface brightness
models on a three-dimensional grid in parameter space of scale-length and
ellipticity.  These models can be discrete Fourier transformed and those
transforms stored for use with all the galaxies.
The choice of a grid of models allows a considerable multiplex gain to
be realised: the models can be pre-generated on that grid and the same
set used for fitting to every galaxy.
\item Estimate the surface brightness profile of the PSF
on the same pixel scale as the models.  Usually this would be done by stacking
images of stars from the region of an image as the galaxies being measured.
The PSF can also be Fourier transformed and stored.  If the PSF varies over
an image or between images, the image may be divided into zones over which
the PSF is approximately invariant, and the Fourier transform of the
PSF for each zone stored separately.  If a mathematical model for the
varying PSF is known this may also be used to generate a smoothly varying
PSF \citep[e.g.][]{rhodes07}.
\item Estimate the rms noise in each pixel from the entire image.
\item Identify a set of nominal galaxy positions to be measured, most
likely from a separate image analysis tool such as SExtractor
\citep{BertinArnouts}.
\item In turn for each galaxy, extract a sub-image centred on that galaxy,
Fourier transform it, and temporarily store the result.
\item For this galaxy, take each possible model in turn,
multiply by the transposed PSF and model transforms to carry out
the cross-correlation, measure the amplitude, width and 
position of the maximum of the resulting
cross-correlation, and hence evaluate the likelihood for this model and galaxy.
Repeat for all models on the grid (or for a subset of models if a more
intelligent maximum-likelihood or MCMC search algorithm is being employed). 
\item Numerically marginalise over the scale-length parameter.  
In the
implementation described here we adopt a uniform prior for the distribution
of galaxy scale-length. This could be replaced by a prior close to the
actual distribution of galaxy sizes, although such a prior would need to be
magnitude-dependent.
\item Discard the extracted data when all models have been explored, and repeat
for the next galaxy.
\end{enumerate}

The result is a grid of likelihood values in  ellipticity parameter space which
thus defines the probability surface $p(\bmath{e})$.  The reduced shear may then
be directly estimated from $\langle \bmath{e} \rangle$, and the uncertainty
in individual $\bmath{e}$ values may be estimated from the width of the likelihood
surface.  

There is a significant multiplex gain obtained by Fourier-transforming the
models, the PSFs and the data and storing the results.  The time-consuming step
then is the cross-correlation, which comprises some multiplications and
a single inverse Fourier transform to obtain the cross-correlation function.
It is this multiplex gain, combined with the elimination of three parameters by
marginalisation, that yields a fast fitting algorithm.

The algorithm is approximate, in the sense that we require the cross-correlation
amplitude to be high enough that $\beta \gg s^2/2b^2$, and also in that we
assume the core of the cross-correlation function can be adequately modelled
as a Gaussian, and we have assumed that the pixel noise is invariant.
This latter constraint may impose a maximum brightness limit on galaxies that
may be fitted, as the pixel noise is not invariant in the case where the galaxy
itself makes a significant contribution to the noise.  For a fixed size of
extracted region around each galaxy, there is also a maximum galaxy size that 
can be adequately measured.  Larger sizes are possible at the expense of greater
computation time.

We note that, in this method, the final PSF that is used is itself a convolution
of PSF components arising from the atmosphere, telescope and the pixel response
of the detector.  We do not need to distinguish the origin of
the final PSF that is used, the method takes a galaxy model and convolves that
with an estimate of the final PSF in order to cross-correlate with the data.
Ultimately however this, and all shape measurement methods, are limited by the extent to
which the sampled data fully encapsulate the information in the sky: the effect
of sampling is to alias spatial frequencies higher than the Nyquist sampling 
frequency.  This affects both the creation of the stacked PSF and the model-fitting
itself. If astronomical observations were band-limited this would not be a
problem, but in reality some aliasing is inevitable.  Poorly sampled observations
should ideally be ``dithered'' in order to reduce such aliasing effects.  

\section{Results}
\label{results}
\subsection{Tests on simulated galaxy images}
The algorithm has been implemented and tested on simulations provided
for the `Shear TEsting Program', STEP \citep{heymans,massey07b}.
Images of galaxies were simulated for the Canada-France-Hawaii telescope
with pixel scale $0.206''$.  The simulations used here to demonstrate
basic shape measurement are those with
an isotropic PSF of FWHM $0.9''$ and zero lensing shear
(tests of shear measurement in the companion paper will cover all the
simulated PSF shapes and shear values).
Simulated galaxies with a mixture of bulge/disc components were used
but all were fitted with a single exponential surface brightness profile.

As here we are testing the Bayesian method, and not our ability to locate
galaxies, we use as input galaxy positions those that were used when
making the simulations.  We also adopt as the prior $\mathcal{P}(\bmath{e})$
the input ellipticity distribution used in the simulations. 

For these tests the size of each subimage was 32 pixels square.  
The choice of subimage size is a compromise between (i) having the subimage
large enough that the galaxy surface brightness distribution is not
unduly truncated and (ii) not allowing the computation time to become
excessively long.  In our initial implementation we have also required
that only a single galaxy should occupy each subimage, thereby eliminating
close pairs.  The choice of 32 pixels for the STEP galaxies ensured that
the subimage was larger than the half-light diameter in every case.
In principle, the subimage size could be a function of galaxy size or
brightness, but this sophistication would introduce some complexity into
the code and has not been tested here.

The likelihood
was evaluated on a cartesian grid in $\bmath{e}$, sampling at intervals of
0.1 out to a maximum axial ratio $a/b=10$.  
The resulting galaxy shapes
were found to be consistent for ellipticity grid intervals less than
0.1 for these STEP galaxies.  The choice of grid interval may need to be
adjusted for different surveys.

The PSF was created by stacking stars from the simulation, 
allowing sub-pixel registration using sinc-function interpolation.
Ultimately any shear measurement survey will be limited by the accuracy to which the
PSF is known.  Systematic PSF errors will of course cause a systematic error in
estimated shear, and if the PSF varies on some angular scale within a survey
this will imprint a signal on that scale on the shear power spectrum.  This concern 
is common to all methods of shape and shear measurement, and we do not specifically
address this problem here.  

\begin{figure}
\resizebox{84mm}{!}{
\includegraphics{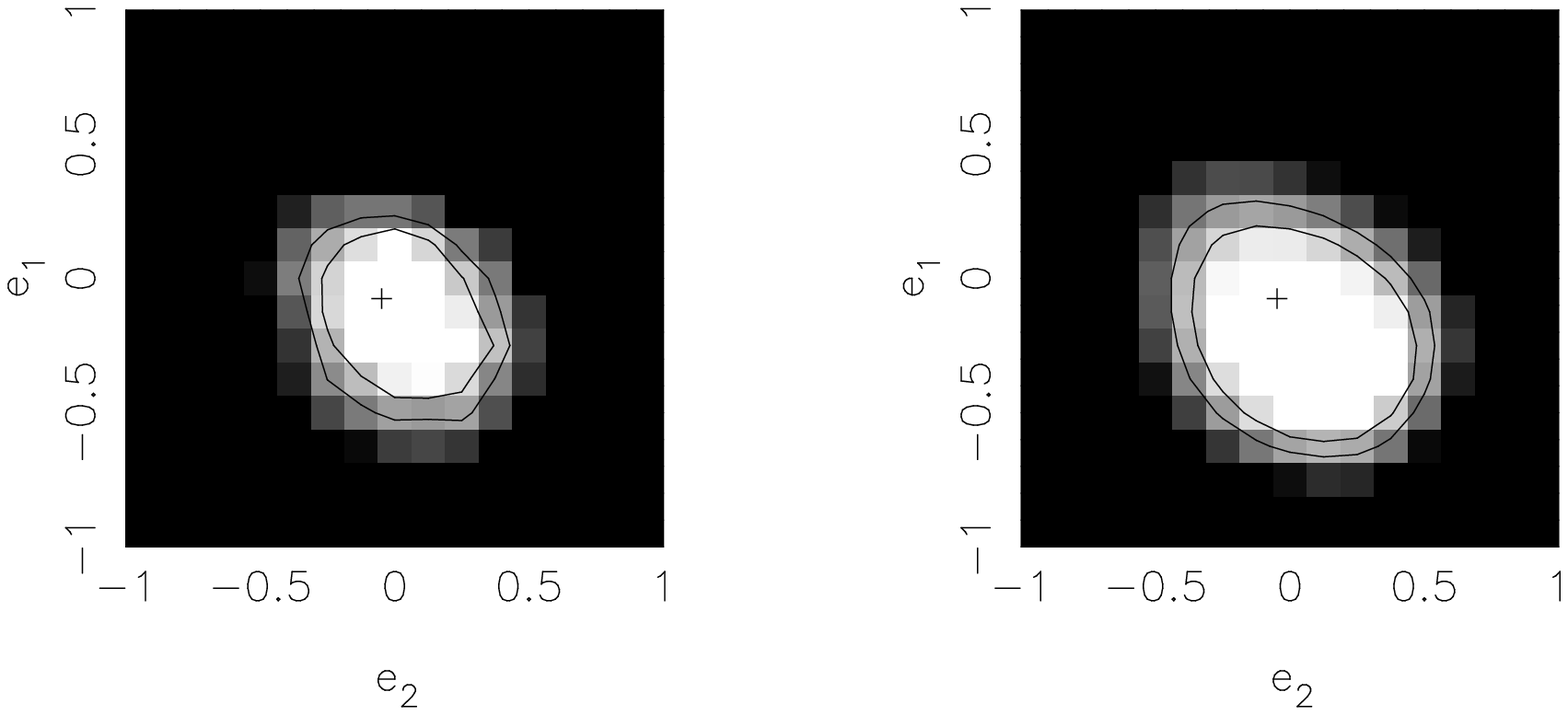}}

\vspace{5mm}

\resizebox{84mm}{!}{
\includegraphics{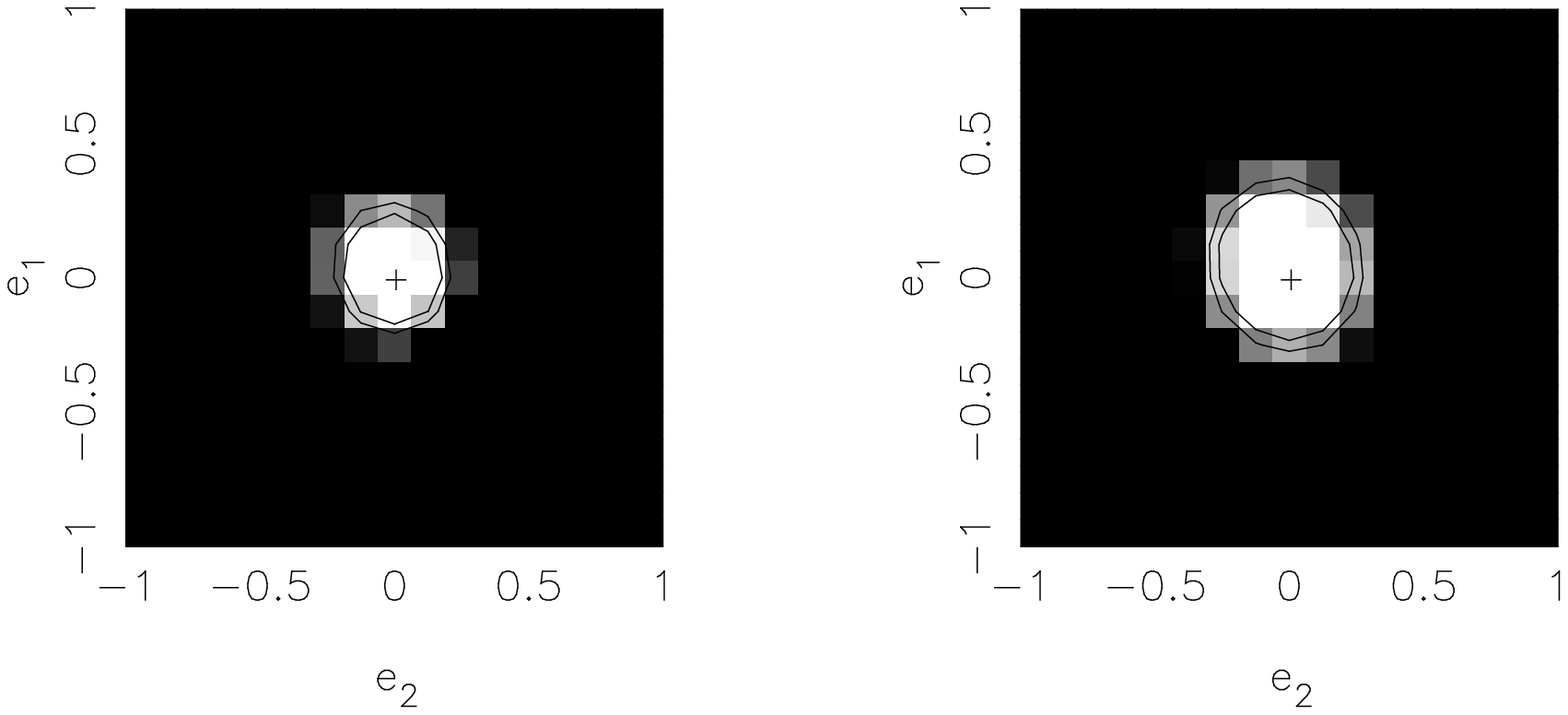}}
\caption
{
Comparison of Bayesian posterior probability
$p(\bmath{e})$ (left) and likelihood $\mathcal{L}(\bmath{e})$ (right) 
surfaces for two individual galaxies.
The grey-scale is logarithmic showing a range of 5
in $\Delta\log\mathcal{L}$ below the maximum value (shown as white) in each case.
The upper panel shows results from fitting
to a magnitude $24.17$ simulated STEP galaxy, 
the lower panel a magnitude $23.15$ galaxy.
Solid lines show the two parameter $1$-$\sigma$ and 
$2$-$\sigma$ contours. The cross shows the input ellipticity value. 
}
\label{fig:likelihood_surfaces}
\end{figure}

\begin{figure}
\resizebox{84mm}{!}{
\includegraphics{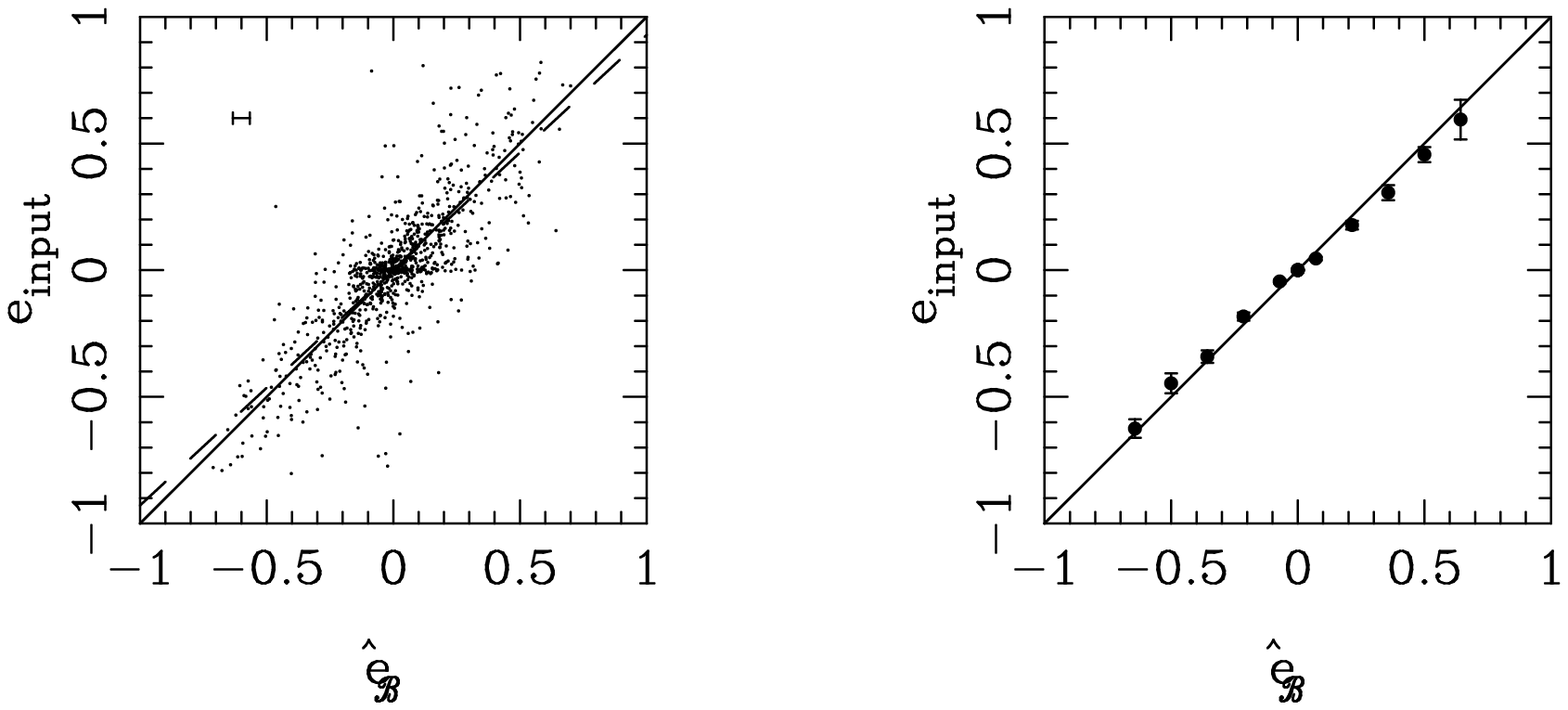}}

\vspace{5mm}

\resizebox{84mm}{!}{
\includegraphics{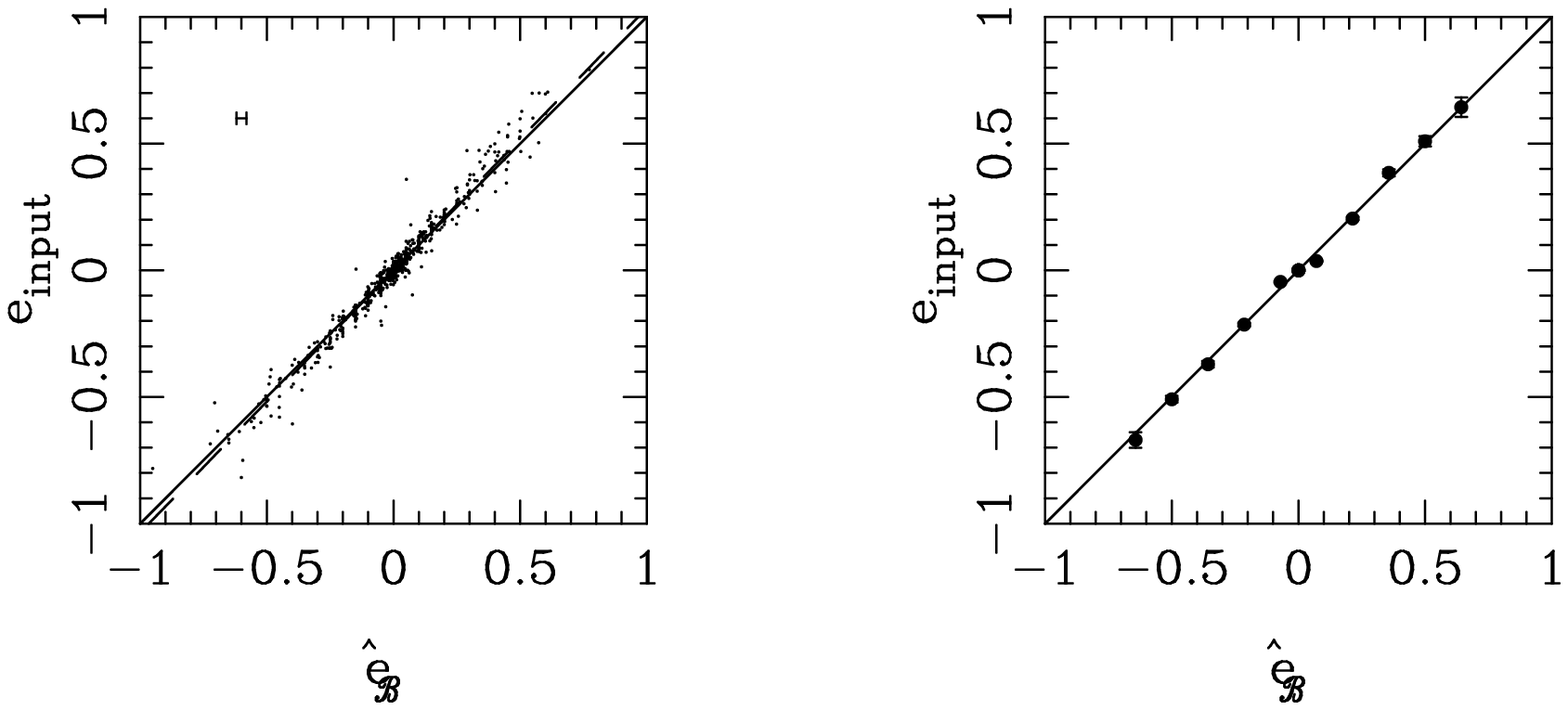}}
\caption
{Tests on the STEP\,1 simulated galaxy sample, as a function of galaxy
apparent magnitude. Each graph shows the expectation value of
the Bayesian estimate of component
$e_1$ (x-axis) plotted against the input value (y-axis).  
Results for component $e_2$ are similar and are not shown.
Left-hand panels
show individual simulated galaxies, right-hand panels show results
binned in intervals of the measured ellipticity.  Two magnitude
ranges are shown, $m > 22$ (upper panels) and $m \le 22$ (lower panels).
The solid lines have a slope of unity, the dashed lines on the left-hand
panels show the least-squares regression of input values on estimated values.
The mean error on individual measured ellipticities is shown
on the left-hand panels.  Vertical error bars on the right-hand panels
indicate the error in the mean input values in each interval of measured values.}
\label{fig:e_plots}
\end{figure}

An assumption of the fast fitting algorithm is that we are fitting to individual galaxies,
and hence close pairs of galaxies cannot be fitted with this algorithm.  In practice
one could identify such close pairs in the data at the galaxy-detection stage, and 
on those galaxies we could use a fitting algorithm that fits multiple components.
In this case the full six parameters per galaxy would need to be fitted, with marginalisation
over uninteresting parameters being carried out post-fitting.  There would still be a 
significant speed advantage to be gained by using the fast fitting algorithm on the
more isolated galaxies however.  In this paper we focus on testing the Bayesian method
and the fast fitting algorithm, and hence in the results presented here we exclude
cases where multiple objects are identified within a single galaxy sub-image.  This procedure
excludes 13\,percent of galaxies in the STEP simulations. Some rejection of close
pairs is required in most other methods of shape estimation also: in future, development
of a fast multiple-component fitting algorithm might allow this constraint to
be relaxed. 
We also test the fit returned by
the fast fitting algorithm to determine whether the fitted centroid of a galaxy is within a
reasonable range of the nominal position, given the prior on the galaxy position: 
this would identify some of the cases of multiple galaxies.  The criterion adopted is
that the fitted galaxy position should lie within $3\sigma$ of the nominal position, where
$\sigma^2$ is the prior position variance, and this excludes 9\,percent of the initial
simulated galaxy sample but no others that are excluded by the `close pairs' criterion.

Fig.\,\ref{fig:likelihood_surfaces} shows the posterior probability surfaces that
result from fitting to two of the simulated galaxies.  For completeness we also
show the likelihood surfaces, which are broader and more biased away from the 
nominal value of ellipticity.

Fig.\,\ref{fig:e_plots} shows the results for each galaxy in the simulations
(only the first component of ellipticity, $e_1$, is shown, similar
results are obtained for $e_2$).
At bright magnitudes there is good correspondence between
input and measured ellipticity values.  The slope appears slightly steeper than
unity, but with a value for the slope of $1.04 \pm 0.08$ the departure from unity
is not very significant.

\begin{figure}
\resizebox{84mm}{!}{
\includegraphics{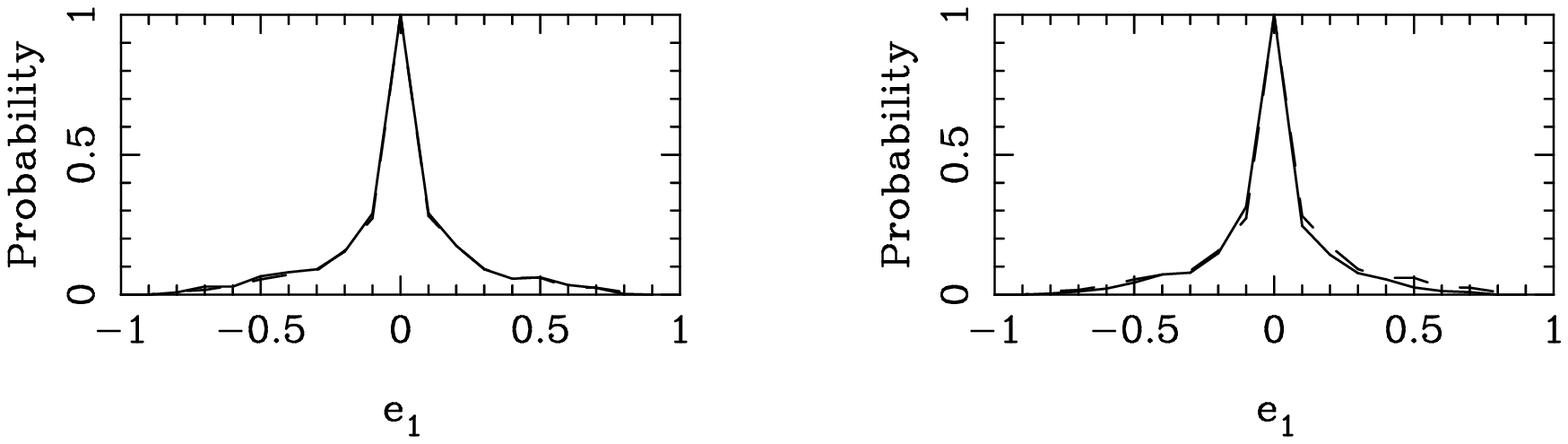}}

\vspace{5mm}

\resizebox{84mm}{!}{
\includegraphics{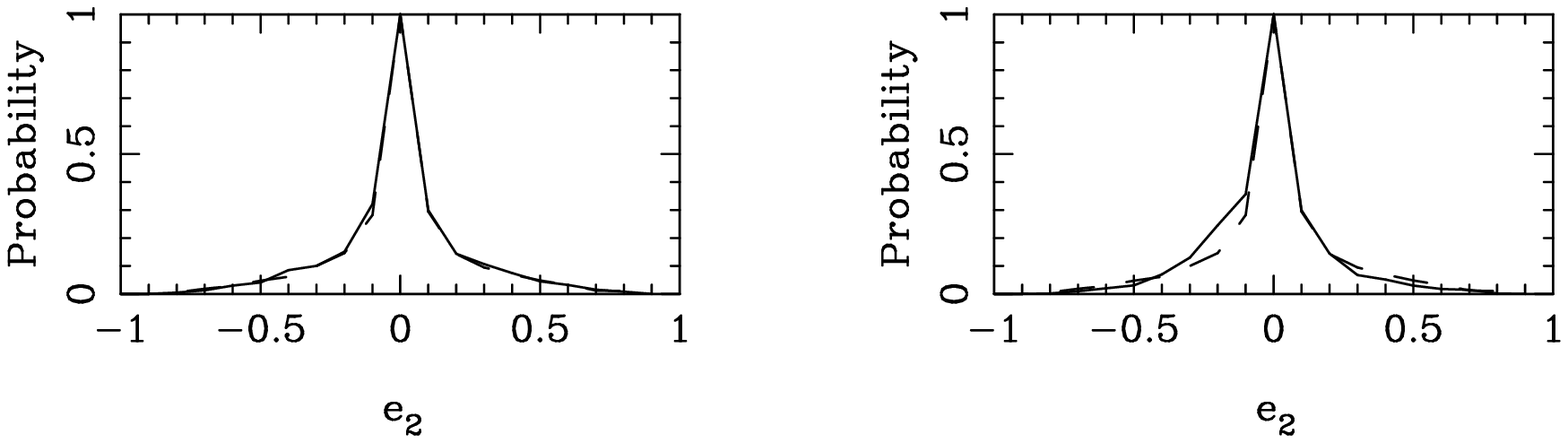}}
\caption
{
The summed posterior probability distribution
of measured ellipticity values $e_1$ (top) and $e_2$ (bottom) as a function of 
apparent magnitude. The prior $\mathcal{P}(\bmath{e})$ is also shown 
for comparison as a dashed line. The magnitude ranges of the simulated galaxies 
are $m\leq 22$ (left panels) and $m>22$ (right panels).
}
\label{fig:e_hist}
\end{figure}

At fainter magnitudes, as the signal-to-noise decreases, an increasing fraction of
galaxies with a given value of the Bayesian measure are drawn from a wider range of
input ellipticities, as expected from the earlier discussion.  The slope of the
relation between input and measured values is again close
to unity, with value $0.93 \pm 0.11$.
At all magnitudes the summed posterior probability 
distribution is a faithful reproduction of
the distribution of the input prior distribution (Fig.\,\ref{fig:e_hist})
as expected from section\,\ref{sampledistribution}.
There is also no detectable correlation between estimated values of
$e_1$ and $e_2$ in this simulated galaxy sample.

We can also investigate the effect of measurement uncertainty in the prior positions
of the galaxies.  A random position uncertainty drawn from a normal distribution
was introduced to each galaxy and the shapes remeasured.  The prior assumed
in the fitting was a normal distribution of rms 3 pixels throughout.
In the results, no change was found
in the slope of Fig.\,\ref{fig:e_plots} for rms position uncertainties as large
as 10 pixels.  The scatter about the mean relation did not change for rms uncertainties
less than 3 pixels and increased by 4 percent for rms uncertainties as large as 10
pixels.  This test indicates that the results are not sensitive to uncertainties
in galaxy position measurement.  
We would recommend that the position prior that is chosen should match the
actual position uncertainty for the faintest galaxies that are reliably fitted.

Tests of the algorithm, again using the full suite of STEP simulations specifically
to measure the shear values recovered, are made in the companion paper (Kitching et al.
in preparation).

\subsection{Speed}
The algorithm has been implemented in the C programming language 
\footnote{The code {\em lensfit} is available on request from the authors: modification to the
data input stages is likely to be required for any particular survey.}
for use on desktop computing systems, with discrete fast Fourier transforms being
supplied by the {\sc FFTW} library \citep{frigo}
\footnote{http://www.fftw.org}.
The computational speed per galaxy obviously depends on the computing
system being used as well as on issues such as the extent to which the 
multiplex advantage of having many galaxies per PSF function can be
exploited.  In the simulations described above, using readily available
2\,GHz desktop PCs in 2007
and evaluating the likelihood surfaces on a grid 
of sampling interval 0.1 in $\bmath{e}$, we found computation times
around 1.0\,s per galaxy, implying that a survey of $10^7$ galaxies
could be analysed in a few months on a single standard desktop PC.
The computation time scales inversely with the square of the sampling
in ellipticity and increases approximately as 
$m^2 \log m$ for subimages of size $m \times m$.

\section{Further considerations}

\subsection{The ellipticity prior}\label{sec:prior}
A number of studies have been made of the distribution of galaxy
ellipticities \citep*[e.g.][]{Lambas, Brainerd, Ebbels,
BernsteinJarvis}.  These studies find a wide variation in distribution
of axial ratios, which appears strongly dependent on apparent
magnitude, presumably largely as a result of the changing mix of
galaxies with brightness and redshift.  The distribution of
ellipticities at the faint magnitudes probed by ongoing and future
weak lensing surveys is even less well-known, and the best estimate
would come from the lensing data itself.  For a sufficiently large
survey the prior estimate could also be allowed to be a function
of galaxy brightness, redshift or colour, if that information were
available.  One way to estimate the
ellipticity prior may be to adopt an iterative approach: evaluate the
summed posterior probability distribution starting from an initial
guess of the prior distribution; then iteratively adjust the assumed
prior until the summed posterior and prior distributions agree.  We
would expect this to be a stable iteration in the absence of sampling
noise, because if the prior is initially assumed distributed to values
that are smaller than are required to explain the data, the next
iteration will adjust the prior to be distributed to large values, and
{\em vice versa}.  Such an approach might however be unstable with
small surveys where sampling noise might be important.

In the case of lensing shear estimation, the ellipticity prior should
also include the shear effect, and should not just be the intrinsic
pre-sheared distribution.  As the shear varies on relatively small
scales, and we are unlikely to have sufficient number of galaxies to
measure accurately the ellipticity distribution in small regions, we
suggest that correct generation of the prior should be to force the
prior to be circularly symmetric, centred on $\langle e \rangle = 0$,
and to be obtained from the large numbers of galaxies that comprise
the full survey.  In this way `false' shear variation arising from
noise on the prior would be avoided, but the resulting shear values
would be slightly biased to low values, in a magnitude-dependent way.
This bias has already been discussed in section\,\ref{shearestimation} and 
a method of correcting for the bias using the shear sensitivity
has been described.

\subsection{Choice of model surface brightness profile}
A key advantage of the model-fitting approach over other methods is that
a surface brightness profile may be chosen that accurately represents
the actual profiles of galaxies.  Two obvious choices of profile are
exponential or de Vaucouleurs.  In fact, it is notoriously difficult to
choose between these profiles when fitting to faint galaxies, so we do not
expect the accuracy of the weak lensing measurement to depend strongly on
which of these profiles is chosen.  Some greater freedom in profile could
be allowed by adding the S\'{e}rsic index as a free parameter, allowing
exponential and de Vaucouleurs models to be treated as special cases of
this generalised profile \citep[e.g.][]{dunlop}, however it is unlikely
that the addition of an extra parameter can be justified on evidence grounds.
A similar consideration is that galaxies generally are composed of bulge
and disc components, which when viewed at an inclined angle may present
differing ellipticities: accurate modelling of galaxies requires both
components to be fitted, but again, for generic shape measurement of faint
galaxies where information at this level of sophistication is not present
in the data, this seems unwarranted.

\subsection{Addition of multiple images or wavebands}
It may be that a weak lensing survey comprises multiple images of the same
region of sky, but taken at different times and hence with differing PSFs,
and possibly in different wavebands.  The latter is likely if broad-band 
optical photometric redshifts are also being estimated from the same data
that is being used for shape measurement.  Clearly one would like to
optimally estimate galaxy shape from the combination of all this data, but
it would not be optimal simply to co-add the data, because of the differing
PSFs in each image.  The model-fitting algorithm described above allows
a natural way to optimally include all the data, since all we need do
is add the likelihoods for the models fitted to each galaxy.  In doing this,
we should take care that the nominal galaxy positions are the same in each
image, so the optimal way to proceed would be to co-add images for the
purpose of detecting and measuring nominal galaxy positions only, and then
fitting each individual image with models convolved with the appropriate
PSF and adding the resulting likelihoods.  Images with a mixture of 
seeing qualities are thus optimally combined for the shape measurement.

\subsection{Weak lensing from radio interferometer data}\label{radio}
It is clear that in large future optical surveys systematic
uncertainties in PSF correction will be a dominating concern,
indeed this is a significant factor in the case for space-based weak
lensing missions.  Ground-based optical PSFs vary  
temporally and very often on spatial scales comparable to
those on which the cosmic shear signal is detectable.  Even
HST lensing studies suffer significantly from PSF variation
\citep{rhodes07,schrabback}.  In principle radio interferometers have
precisely known PSFs, being determined by
the antenna positions (note that full 3D knowledge of antenna positions
is required, to allow for curvature of the Earth and natural height
variations).  The PSF varies with hour angle and declination,
but in a completely deterministic way.  Other effects such as
bandwidth and sampling-time smearing can also be precisely computed
and incorporated into the shape measurement process \citep{Chang04}.
Because interferometer measurement are made in the Fourier domain,
and because the noise also originates in that domain (being associated
with individual antennas) it makes sense to measure galaxy shapes
in Fourier space (in the image plane the noise is correlated between
pixels, effectively being also convolved with the PSF).
\citet{ChangRefregier} and \citet{Chang04} have already
shown how a shapelets \citep{refregier, RefregierBacon} 
based approach can be extended to the Fourier domain.  The Bayesian algorithm
presented in this paper already operates in the Fourier domain, so it
should be easily adapted for radio interferometer data, which will be
particularly relevant for future deep radio surveys such as those proposed
for the Square Kilometer Array.

\section{Conclusions}

We have argued that a model-fitting approach to galaxy shape
measurement should provide an optimum approach to shape measurement
for large weak-lensing surveys, with the advantages that the
signal-to-noise of the shape measurement should be optimised and
random measurement errors can be estimated.  We have further argued
that a Bayesian estimation process allows unbiased shape estimation to
be made, although even in a realistic implementation of a Bayesian
method there is a bias in recovered shear values introduced by the
presence of the prior probability distribution.  This bias may be
calculated from the measured likelihood surfaces, however, and in this
paper we have spent some time discussing the calculation of the `shear
sensitivity'.  Overall this approach to shape measurement should
provide a framework for shear measurement that does not need external
calibration by comparison with simulations.
 
A traditional disadvantage of model-fitting is that it may be
computationally time-consuming, and in this paper we present a fast
algorithm for measuring the shapes of individual galaxies.  The
algorithm makes use of analytic marginalisation over surface
brightness amplitude, and by working in Fourier space enables rapid
marginalisation over galaxy position.  The algorithm has been tested
and has an adequate speed on current generations of computers for use
with large ongoing and planned weak-lensing surveys.  Close pairs of
galaxies are not treated by the algorithm, but provided such close
pairs can be identified in the data a separate fitting process may be
applied to those.

The Bayesian method and fast fitting algorithm have been tested on
simulated galaxies created for the Shear TEsting Program (STEP:
\citealt{heymans, massey07b}) and promising results on the measurement
of individual galaxy ellipticities have been obtained.  A companion
paper (Kitching et al. in preparation) will test the measurement of
weak lensing shear in the STEP simulations.

\vspace*{5mm}
\noindent
{\large \bf ACKNOWLEDGEMENTS}\\ 
CH acknowledges the support of a European Commission Programme $6^{\rm
th}$ framework Marie Curie Outgoing International Fellowship under
contract M01F-CT-2006-21891, and a CITA National Fellowship.  LVW is
supported by NSERC, CIfAR and CFI. TDK is supported by the Science and
Technology Facilities Council, research grant number E001114.


\label{lastpage}


\begin{thebibliography}{99}
\bibitem[\protect\citeauthoryear{Bacon \& Taylor}{2003}]{BaconTaylor}
Bacon D.J., Taylor A.N., MNRAS 344 (2003) 1307
\bibitem[\protect\citeauthoryear{Bardeau et al.}{2005}]{bardeau05}
Bardeau S., Kneib J.-P., Czoske O., Soucail G., Smail I., Ebeling H., Smith G.P., 2005, A\&A, 434, 433
\bibitem[\protect\citeauthoryear{Bardeau et al.}{2007}]{bardeau07}
Bardeau S., Soucail G., Kneib J.-P., Czoske O., Ebeling H., Hudelot P., Smail I., Smith G.P., 2007, A\&A, 470, 449
\bibitem[\protect\citeauthoryear{Benjamin et al.}{2007}]{benjamin07}
Benjamin J. et al., 2007, MNRAS submitted, astro-ph/0703570
\bibitem[\protect\citeauthoryear{Bernstein \& Jarvis}{2002}]{BernsteinJarvis}
Bernstein G.M., Jarvis M., 2002, AJ, 123, 583
\bibitem[\protect\citeauthoryear{Bertin \& Arnouts}{1996}]{BertinArnouts}
Bertin E., Arnouts S., 1996, A\&AS, 117, 393
\bibitem[\protect\citeauthoryear{Brainerd, Blandford \& Smail}{Brainerd et al.}{1996}]{Brainerd}
Brainerd T.G., Blandford R.D., Smail I., 1996, ApJ, 466, 623
\bibitem[\protect\citeauthoryear{Bridle et al.}{2002}]{bridle}
Bridle S., Kneib J.-P., Bardeau S., Gull S., 2002, {\em in}
The shapes of galaxies and their dark halos, Proceedings of the Yale Cosmology Workshop "The Shapes of Galaxies and Their Dark Matter Halos", New Haven, Connecticut, USA, 28-30 May 2001. Edited by Priyamvada Natarajan. Singapore: World Scientific,
\bibitem[\protect\citeauthoryear{Chang \& Refregier}{2002}]{ChangRefregier}
Chang T.-C., Refregier A., 2002, ApJ, 570, 447
\bibitem[\protect\citeauthoryear{Chang, Refregier \& Helfand}{Chang et al.}{2004}]{Chang04}
Chang T.-C., Refregier A., Helfand D.J., 2004, ApJ, 617, 794
\bibitem[\protect\citeauthoryear{Dunlop et al.}{2003}]{dunlop}
Dunlop J.S., McLure R.J., Kukula M.J., Baum S.A., O'Dea C.P., Hughes D.H., 2003,
MNRAS, 340, 1095
\bibitem[\protect\citeauthoryear{Ebbels, Kneib \& Ellis}{Ebbels et al.}{1999}]{Ebbels}
Ebbels T., Kneib J.-P., Ellis R.S., 1999, {\em in}
Cosmological Parameters and the Evolution of the Universe. Edited by Katsuhiko Sato. Publisher: Dordrecht, Boston: Kluwer Academic, 1999. ("Proceedings of the 183rd symposium of the International Astronomical Union held in Kyoto, Japan, August 18-22, 1997", p. 247
\bibitem[\protect\citeauthoryear{Edmondson, Miller \& Wolf}{Edmondson et al.}{2006}]{edmondson}
Edmondson, E.M., Miller, L., Wolf, C., 2006, MNRAS, 371, 1639
\bibitem[\protect\citeauthoryear{Frigo \& Stevens}{2005}]{frigo}
Frigo M., Johnson S.G. 2005, Proceedings of the IEEE, 93, 216
\bibitem[\protect\citeauthoryear{Heavens}{2003}]{Heavens}
Heavens A., 2003, MNRAS, 343, 1327
\bibitem[\protect\citeauthoryear{Heavens, Kitching \& Taylor}{Heavens et al.}{2006}]{Heavens06}
Heavens A.F., Kitching T.D., Taylor A.N., 2006, MNRAS, 373, 105
\bibitem[\protect\citeauthoryear{Heymans et al.}{2006}]{heymans}
Heymans C. et al. 2006, MNRAS, 368, 1323
\bibitem[\protect\citeauthoryear{Hu}{1999}]{Hu}
Hu W., 1999, ApJ, 522, 21
\bibitem[\protect\citeauthoryear{Jing}{2006}]{jing}
Jing Y.P., Zhang P., Lin W.P., Gao L., Springer V., 2006, ApJ, 640, L119
\bibitem[\protect\citeauthoryear{Kaiser, Squires \& Broadhurst}{Kaiser et al.}{1995}]{ksb}
Kaiser N., Squires G., Broadhurst T., 1995, ApJ, 449, 460
\bibitem[\protect\citeauthoryear{Kaiser}{2000}]{kaiser00}
Kaiser N., 2000, ApJ, 537, 555
\bibitem[\protect\citeauthoryear{Kitching et al.}{2007}]{Kitching}
Kitching T.D., Heavens A.F., Taylor A.N., Brown M.L., Meisenheimer K., Wolf C., 
Gray M.E., Bacon D.J., 2007, MNRAS, in press
\bibitem[\protect\citeauthoryear{Kneib et al.}{2003}]{kneib}
Kneib J.-P., Hudelot P., Ellis R.S., Treu T., Smith G.P., Marshall P., Czoske O., Smail I., Natarajan P., 2003, ApJ, 598, 804
\bibitem[\protect\citeauthoryear{Kuijken}{1999}]{kuijken99}
Kuijken K., 1999, A\&A, 352, 355
\bibitem[\protect\citeauthoryear{Kuijken}{2006}]{kuijken06}
Kuijken K., 2006, A\&A, 456, 827
\bibitem[\protect\citeauthoryear{Lambas, Maddox \& Loveday}{Lambas et al.}{1992}]{Lambas}
Lambas D.G., Maddox S.J., Loveday J., MNRAS, 258, 404
\bibitem[\protect\citeauthoryear{Luppino \& Kaiser}{1997}]{Luppino}
Luppino, G.A., Kaiser, N., 1997, ApJ, 475, 20
\bibitem[\protect\citeauthoryear{Massey et al.}{2007a}]{massey07a}
Massey R. et al., 2007a, ApJ in press, arXiv:astro-ph/0701480
\bibitem[\protect\citeauthoryear{Massey et al.}{2007b}]{massey07b}
Massey R. et al., 2007b, MNRAS, 376, 13
\bibitem[\protect\citeauthoryear{Peng et al.}{2003}]{peng}
Peng C.Y., Ho L.C., Impey C.D., Rix H.-W., 2003, AJ, 124, 266
\bibitem[\protect\citeauthoryear{Refregier}{2003}]{refregier}
Refregier A., 2003, MNRAS, 338, 35
\bibitem[\protect\citeauthoryear{Refregier \& Bacon}{2003}]{RefregierBacon}
Refregier A., Bacon, D., 2003, MNRAS, 338, 48
\bibitem[\protect\citeauthoryear{Rhodes et al.}{2007}]{rhodes07}
Rhodes J.D., Massey R., Albert J., Collins N., Ellis R.S., Heymans C., 
Gardner J.P., Kneib J.-P., Koekemoer A., Leauthaud A., Mellier Y., 
Refregier A., Taylor J.E., Van Waerbeke, L., 2007, ApJ in press
\bibitem[\protect\citeauthoryear{Schrabback et al.}{2007}]{schrabback}
Schrabback T., Erben T., Simon P., Miralles J.-M., Schneider P., Heymans C., Eifler T., Fosbury R.A.E., Freudling W., Hetterscheidt M., Hildebrandt H., Pirzkal N.,
2007, A\&A, 468, 823
\bibitem[\protect\citeauthoryear{Seitz \& Schneider}{1997}]{SeitzSchneider}
Seitz C., Schneider P., 1997, A\&A, 318, 687
\bibitem[\protect\citeauthoryear{Schramm \& Kayser}{1995}]{SchrammKayser}
Schramm T., Kayser R., 1995, A\&A, 299, 1
\bibitem[\protect\citeauthoryear{Spergel et al.}{2007}]{spergel07}
Spergel, D. et al. 2007, astro-ph/0603449
\bibitem[\protect\citeauthoryear{Taylor et al.}{2007}]{Taylor07}
Taylor A.N., Kitching T.D., Bacon D.J., Heavens A.F., 2007, MNRAS, 374, 1377
\bibitem[\protect\citeauthoryear{Tyson, Wenk \& Valdes}{Tyson et al.}{1990}]{tyson}
Tyson J.A., Wenk R.A., Valdes F., 1990, ApJ, 349, L1
\bibitem[\protect\citeauthoryear{White}{2004}]{white}
White M., 2004, Astropart. Phys., 22, 211
\bibitem[\protect\citeauthoryear{Zhan \& Knox}{2004}]{zhan}
Zhan H., Knox L., 2004, ApJ, 616, L75
\end{thebibliography}
\end{document}